\documentclass[superscriptaddress,twocolumn,aps,prb,floatfix,showpacs,amsmath,amssymb,reprint]{revtex4}
\usepackage{epsfig}
\usepackage{dcolumn}
\usepackage{bm}
\begin{document}

\title{Conductance noise in an out-of-equilibrium two-dimensional electron system}

\author{Ping V. Lin}
\email{lin@magnet.fsu.edu}\affiliation{National High Magnetic Field Laboratory, Florida State
University, Tallahassee, Florida 32310, USA}
\author{Xiaoyan Shi}
\email{shi@magnet.fsu.edu}\affiliation{National High Magnetic Field Laboratory, Florida State University, Tallahassee, Florida 32310, USA}
\affiliation{Department of Physics, Florida State University, Tallahassee, Florida 32306, USA}
\author{J. Jaroszynski}
\email{jaroszy@magnet.fsu.edu}\affiliation{National High Magnetic Field Laboratory, Florida State
University, Tallahassee, Florida 32310, USA}
\author{Dragana Popovi\'c}
\email{dragana@magnet.fsu.edu}\affiliation{National High Magnetic Field Laboratory, Florida State University, Tallahassee, Florida 32310, USA}
\affiliation{Department of Physics, Florida State University, Tallahassee, Florida 32306, USA}

\date{\today}

\begin{abstract}
A study of the conductance noise in a two-dimensional electron system (2DES) in Si at low temperatures ($T$) reveals the onset of large, non-Gaussian noise after cooling from an equilibrium state at a high $T$ with a fixed carrier density $n_s$.  This behavior, which signifies the falling out of equilibrium of the 2DES as $T\rightarrow 0$, is observed for $n_s<n_g$ ($n_g$ -- glass transition density).  A protocol where density is changed by a small value $\Delta n_s$ at low $T$ produces the same results for the noise power spectra.  However, a detailed analysis of the non-Gaussian probability density functions (PDFs) of the fluctuations reveals that $\Delta n_s$ has a qualitatively different and more dramatic effect than $\Delta T$, suggesting that $\Delta n_s$ induces strong changes in the free energy landscape of the system as a result of Coulomb interactions.  The results from a third, waiting-time ($t_w$) protocol, where $n_s$ is changed temporarily during $t_w$ by a large amount, demonstrate that non-Gaussian PDFs exhibit history dependence and an evolution towards a Gaussian distribution as the system ages and slowly approaches equilibrium.  By calculating the power spectra and higher-order statistics for the noise measured over a wide range of the applied voltage bias, it is established that the non-Gaussian noise is observed in the regime of Ohmic or linear response, i.e. that it is not caused by the applied bias.
\end{abstract}

%

\maketitle

\section{INTRODUCTION}

Many novel materials find themselves close to the metal-insulator transition (MIT), where strong electron-electron interactions and disorder are believed to give rise to a variety of complex phenomena.\cite{Vlad-bookintro}  Near the MIT, these two effects are usually comparable in magnitude, so that their competition is expected to lead to glassy behavior of electrons, in analogy with other frustrated systems.\cite{Eduardo-Vlad_review,Vlad-bookchapter}  A common denominator for all glasses is the existence of a complex or ``rugged'' free energy landscape (FEL), consisting of a large number of metastable states, separated by barriers of different heights.  This results in phenomena such as slow, nonexponential relaxations, divergence of the equilibration time, and breaking of ergodicity, i.e. the inability of the system to equilibrate on experimental time scales.  Therefore, such out-of-equilibrium systems also exhibit aging effects,\cite{Struik,aging} where the response to an external excitation (i.e. relaxation) depends on the system history in addition to the time $t$.  A detailed analysis of temporal fluctuations (noise) of the relevant observables yields complementary information on configurational rearrangements or transitions between metastable states.  Non-Gaussian distributions of various observables in glassy systems have been reported,\cite{hetero-review} reflecting the presence of large, \textit{collective} rearrangements.

Most experimental studies of charge or Coulomb glasses have focused on situations where electrons are strongly localized due to disorder, far from the MIT.\cite{Amir-eglass_review}  In recent years, however, a two-dimensional electron system (2DES) in Si metal-oxide-semiconductor field-effect transistors (MOSFETs) has emerged as an excellent model system for studying glassy or out-of-equilibrium charge dynamics near the MIT.\cite{DP-bookchapter}  In particular, studies of both relaxations and noise on very disordered samples have established that the 2DES in Si exhibits all the main manifestations of glassiness: slow, correlated dynamics (non-Gaussian noise),\cite{Snezana2002,Jan2002} nonexponential relaxations,\cite{Jan2006} diverging equilibration time (as temperature $T\rightarrow 0$),\cite{Jan2006} aging and memory.\cite{Jan2007-1,Jan2007-2,Jan2009}  Glassiness is observed for all carrier densities $n_s<n_g$, where $n_g>n_c$ ($n_c$ -- the critical density for the MIT), thus giving rise to an intermediate glassy, ``bad metal'' phase ($k_{F}l <1$; $k_{F}$ -- Fermi wave vector, $l$ -- mean free path).  In a 2DES with a relatively low amount of disorder, noise studies show that the intermediate phase becomes vanishingly small, i.e. $n_g\gtrsim n_c$.\cite{Jan2002,Jan2004}  These observations are consistent with predictions of the theoretical models that describe the MIT as a Mott transition with disorder.\cite{Darko}  The results presented here were obtained on samples with a relatively large amount of disorder, where the intermediate ($n_c<n_s<n_g$) glassy phase is more pronounced, as mentioned above.  We note that aging studies have found\cite{Jan2007-2,Jan2009} that an abrupt change in the nature of the glassy phase occurs at $n_c$, i.e. at the 2D MIT, signifying the intrinsic glassiness of the 2DES.

Glassy dynamics in a 2DES in Si has been studied using several different experimental protocols, all of which involved rapid changes of $n_s$ (``$\Delta n_s$ protocol'').  For example, a large, rapid (within 1~s) change of $n_s$ at a low enough $T$ results in a nonexponential relaxation of conductivity $\sigma(t)$ towards a new equilibrium value.\cite{Jan2006}  Small changes of $n_s$ at low $T$ do not produce any observable relaxations,\cite{Jan2007-2} but have led to the observations of non-Gaussian fluctuations in $\sigma(t)$.\cite{Snezana2002,Jan2002,Jan2004}  On the other hand, in other types of glasses (e.g. structural, spin), studies of aging processes and other glassy properties are typically carried out following a thermal quench from an equilibrium state at a high $T$ to a nonequilibrium state at a low $T$.  The analysis of the relaxations following such a thermal quench, however, is greatly complicated by their nontrivial dependence on the cooling time and, hence, a fairly fast cooling process is essential (see, e.g., Ref.~\onlinecite{SG-fullaging}).  It is of fundamental interest to establish whether the charge glass realized in a 2DES in Si also falls out of equilibrium as $T$ is reduced, in common to other types of glasses.

Here we investigate the charge dynamics in a 2DES in Si at low $n_s$, near the MIT, after cooling from an equilibrium state at a high $T$ with a fixed $n_s$ (``$\Delta T$ protocol'').  Since it is not possible to achieve rapid cooling in our experimental set-up (the shortest cooling time is about 30 minutes), we focus on the resulting fluctuations (noise) in $\sigma(t)$.  After describing the samples and measurement techniques in more detail (Sec.~\ref{expt}), we present the analysis of the noise power spectra obtained using a $\Delta T$ protocol (Sec.~\ref{noisepower}).  The results demonstrate the glassy arrest of the charge dynamics for $n_s<n_g$, where the glass transition density $n_g$ is the same as that found in noise studies that employed $\Delta n_s$ protocols.\cite{Snezana2002,Jan2002}

The power spectrum, being the Fourier transform of a temporal correlation function, only provides information about the second moment of the fluctuating quantity.  Additional information can be extracted from the full probability distribution of the fluctuations.  Therefore, in Sec.~\ref{pdf}, we compare the probability density functions (PDFs) obtained in the two protocols, $\Delta n_s$ and $\Delta T$.  While in both cases the PDFs are non-Gaussian for $n_s<n_g$, the change of the carrier density has a qualitatively different and more dramatic effect than $\Delta T$.  In particular, the results suggest that $\Delta n_s$ induces strong changes in the free energy landscape of the system, which should have important implications for theoretical modeling of the glassy dynamics in a 2DES.

For a $\Delta T$ protocol, the power spectra and the second spectra\cite{Weissman88,Weissman92,Weissman93} of the noise in the glassy regime have also been investigated for different values of the excitation voltage $V_{\mathrm{exc}}$ used in the measurements of $\sigma$.  In indium oxide, another realization of a Coulomb glass, the non-Gaussian noise was attributed to a non-linear mechanism caused by the high applied bias.\cite{Zvi-nonlinear}  As shown in Sec.~\ref{exc}, in the 2DES in Si the large, non-Gaussian noise does not depend on the applied bias over more than two orders of magnitude variation of $V_{\mathrm{exc}}$, providing additional evidence that it reflects the intrinsic glassiness of the 2DES.  The main results are summarized and discussed further in Sec.~\ref{concl}.

\section{EXPERIMENT}
\label{expt}

\subsection{Samples and measurement techniques}

Measurements were carried out on rectangular n-channel (100)-Si MOSFETs with a relatively large amount of disorder.  Similar to previous studies,\cite{Snezana2002,Jan2006,Jan2007-1,Jan2007-2,Jan2009} the back-gate bias of $-2$~V was applied to maximize the 4.2~K peak mobility to $\sim$0.06~m$^2$/Vs.  The sample dimensions $L \times W$ ($L$--length, $W$--width) were $1 \times 90$~$\mu$m$^2$ for samples A and A1, and $2 \times 50$~$\mu$m$^2$ for samples B and B1.  The devices were fabricated with poly-Si gates, self-aligned ion-implanted contacts, substrate doping $N_a \sim 2 \times 10^{17}$~cm$^{-3}$, oxide charge $N_{\mathrm{ox}} \sim 1 \times 10^{11}$~cm$^{-2}$, and oxide thickness $d_{\mathrm{ox}}=50$~nm.  All four samples exhibited the same behavior, and some of them were, in fact, used in previous studies (e.g. A and B in Refs.~\onlinecite{Jan2006,Jan2007-1,Jan2007-2,Jan2009}, A1 in Ref.~\onlinecite{Snezana2002}).

The conductivity $\sigma$ was measured using a standard two-probe ac technique (typically at $\sim 11$ or 13~Hz) with an ITHACO 1211 current preamplifier, and either a PAR124A or a SR7265 lock-in amplifier.  The contact resistances and the contact noise were determined to be negligible in these samples.\cite{Snezana2002}
Unless noted otherwise (see Sec.~\ref{exc}), the excitation voltage $V_{\mathrm{exc}}$ was kept constant and low enough ($2-10~\mu$V) to ensure that the conduction was Ohmic.  The low-frequency of $V_{\mathrm{exc}}$ was limited by the low cut-off frequency of RC filters, which were used to reduce external electromagnetic noise, as well as by high resistance of the samples.  A precision dc voltage standard (EDC MV116J) was used to apply the gate voltage $V_g$, which controls $n_s$: $n_s(10^{11}$cm$^{-2})=4.31(V_g[$V$]-6.3)$.  Measurements of current ($I$) fluctuations as low as $10^{-13}$~A were performed both in a dilution refrigerator and in a $^3$He system (base $T=0.24$~K).  Relatively small fluctuations of $T$, $V_g$, and $V_{\mathrm{exc}}$ were ruled out as possible sources of the measured noise, since no correlation was found between them and the fluctuations of $I$.  The background noise was measured for all $n_s$ and $T$ by setting $V_{\mathrm{exc}}$ to zero.  It was always independent of frequency (i.e. white) and usually several orders of magnitude smaller than the sample noise.

\subsection{Temperature dependence of the time-averaged conductivity}

\begin{figure}
\includegraphics[width=8.1cm]{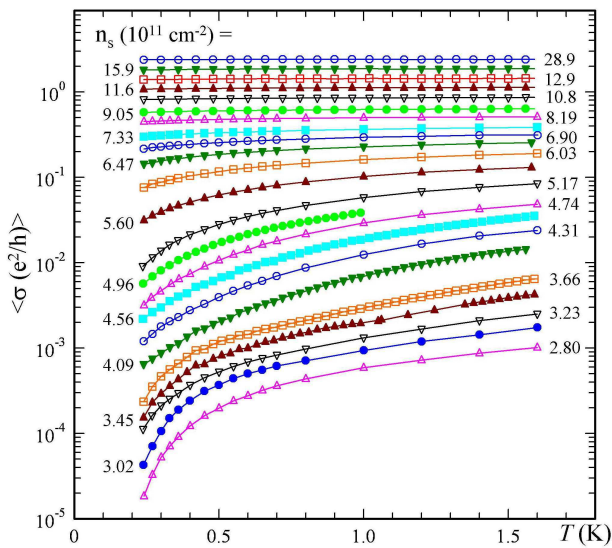}
\caption {(Color online) Sample B.  Conductivity $\langle\sigma\rangle$ \textit{vs.} $T$ for different $n_s$, as shown.  $n_s$ was varied at high $T\approx 10-12$~K.  Solid lines guide the eye. }\label{fig:RvT}
\end{figure}

Figure~\ref{fig:RvT} shows the time-averaged (see Sec.~\ref{noisepower}) conductivity $\langle \sigma \rangle$ as a function of $T$ for
%
%
different $n_s$ for one of the samples.  The density was varied at $T\approx 10-12$~K, where the 2DES is in a thermal equilibrium.\cite{Jan2006}  The critical density $n_c$ is determined from the fits to $\langle\sigma(n_s, T)\rangle$ on both metallic and insulating sides of the MIT, as discussed in detail elsewhere.\cite{Snezana2002,Jan2002,Jan2004,Jan2009}  In particular, in the insulating regime, the transport is thermally activated, whereas in the intermediate, $n_c<n_s<n_g$ glassy phase, $\langle\sigma(n_s, T)\rangle$ acquires a very specific, non-Fermi liquid $T^{3/2}$ correction.  In all samples, $n_c=(4.5\pm 0.4)\times 10^{11}$~cm$^{-2}$, and the glass transition density determined from $\Delta n_s$ protocols\cite{Snezana2002,Jan2006,Jan2007-1,Jan2007-2,Jan2009} was $n_g=(7.5\pm 0.3)\times 10^{11}$~cm$^{-2}$.  The error bars account for sample-to-sample variations.

\section{Conductance noise from a $\Delta T$ protocol}
\label{noisepower}

\begin{figure}
\includegraphics[width=7.1cm]{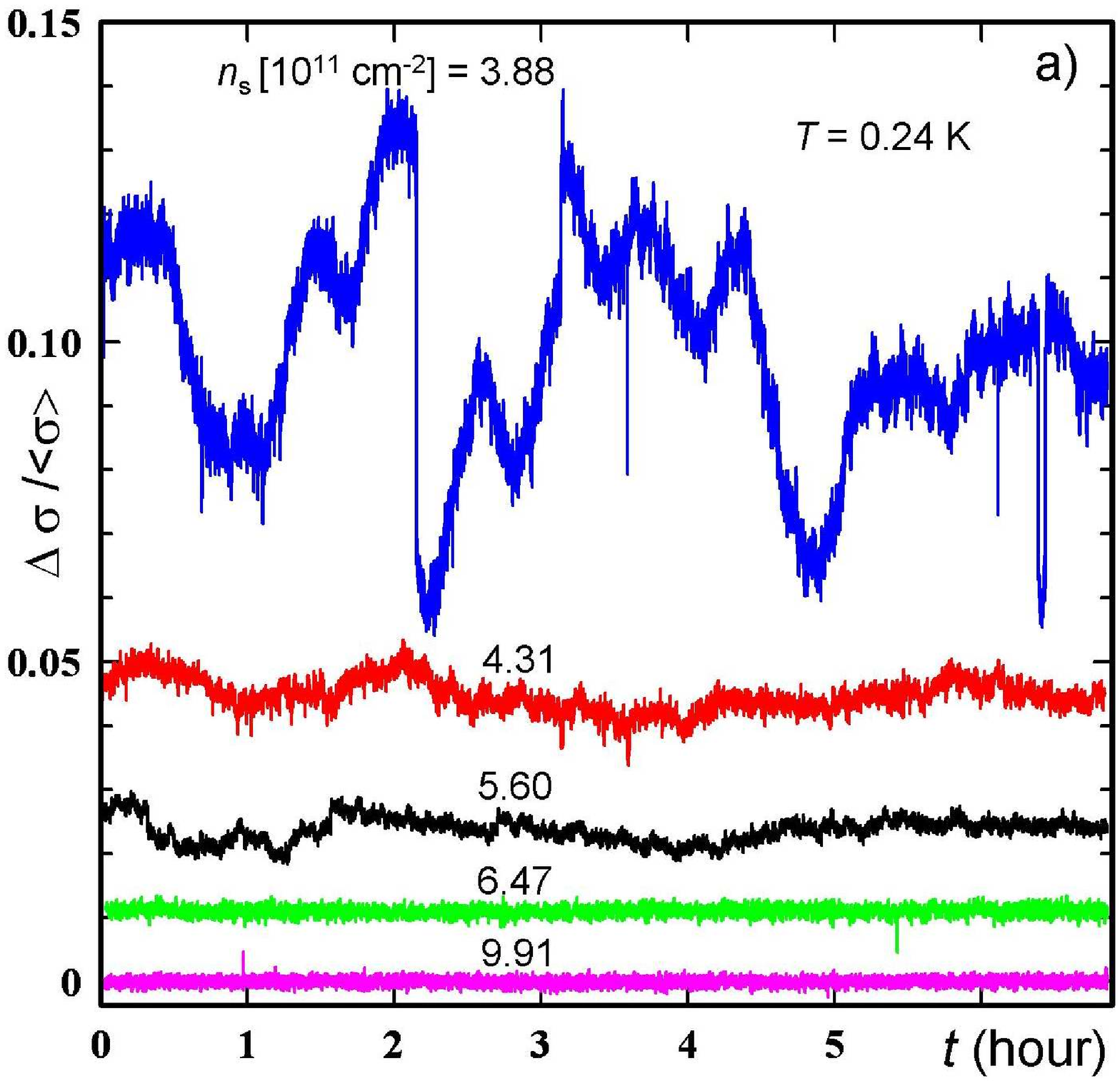}\\
\includegraphics[width=7.1cm]{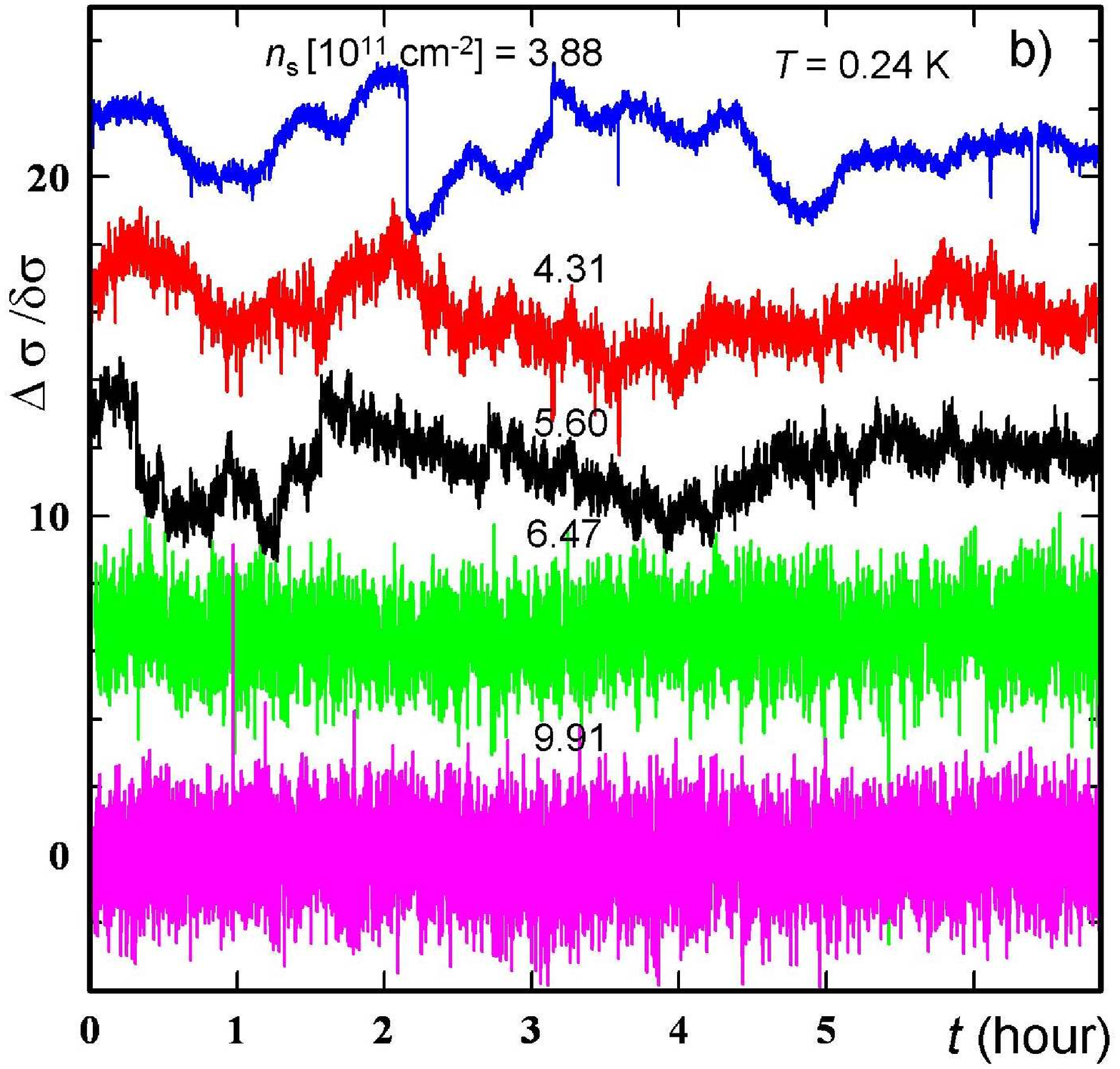}
\caption {(Color online)  Sample B.  (a) Relative fluctuations $\Delta\sigma(t)/\langle\sigma\rangle=(\sigma(t)-\langle\sigma\rangle)/\langle\sigma\rangle$ vs time for several $n_s$ shown on the graph (top to bottom) at $T=0.24$~K. The averaging of $\sigma(t)$ was done over a time interval shown on the plot.  Traces are shifted for clarity.  $n_s$ was changed at $T\approx 10-12$~K and the cooling time was about $10^4$~s.  (b) In order to make the change in the character of the noise with $n_s$ more apparent, the same fluctuations $\Delta\sigma(t)=(\sigma(t)-\langle\sigma\rangle)$ are shown normalized by $\delta\sigma$, where $\delta\sigma=\langle(\sigma-\langle\sigma\rangle)^{2}\rangle^{1/2}$.
}\label{fig:fig2}
\end{figure}

In a $\Delta T$ protocol, each noise measurement starts by fixing the carrier density $n_s$ at $T\approx 10-12$~K.  The sample is then cooled to the measurement temperature.  The cooling times varied from 30 minutes to 10 hours, with no effect on the noise results.  In addition, such slow cooling did not result in any visible relaxations, at least on our experimental time scales. Typical data are presented in Fig.~\ref{fig:fig2}(a), which shows the relative fluctuations $\Delta\sigma(t)/\langle\sigma\rangle$ vs time for several $n_s$ measured at $T=0.24$~K.
%
%
It is obvious that, by decreasing $n_s$, the amplitude of the fluctuations increases dramatically.  Even more striking is the change in the character of the noise: at low $n_s$, fluctuations occur on many different time scales, including slow changes over periods of several hours; at high $n_s$, on the other hand, the variance of the noise no longer varies with time.  These features are more apparent in Fig.~\ref{fig:fig2}(b), where we choose to plot the same $\Delta\sigma(t)=(\sigma(t)-\langle\sigma\rangle)$ normalized by $\delta\sigma$, the square root of the variance of the fluctuations.  The increase in the noise magnitude with decreasing $n_s$ is clearly accompanied by the onset of non-Gaussian behavior, similar to the noise results obtained from a $\Delta n_s$ protocol.\cite{Snezana2002,Jan2002,Jan2004}
%
\begin{figure}
\includegraphics[width=7.1cm]{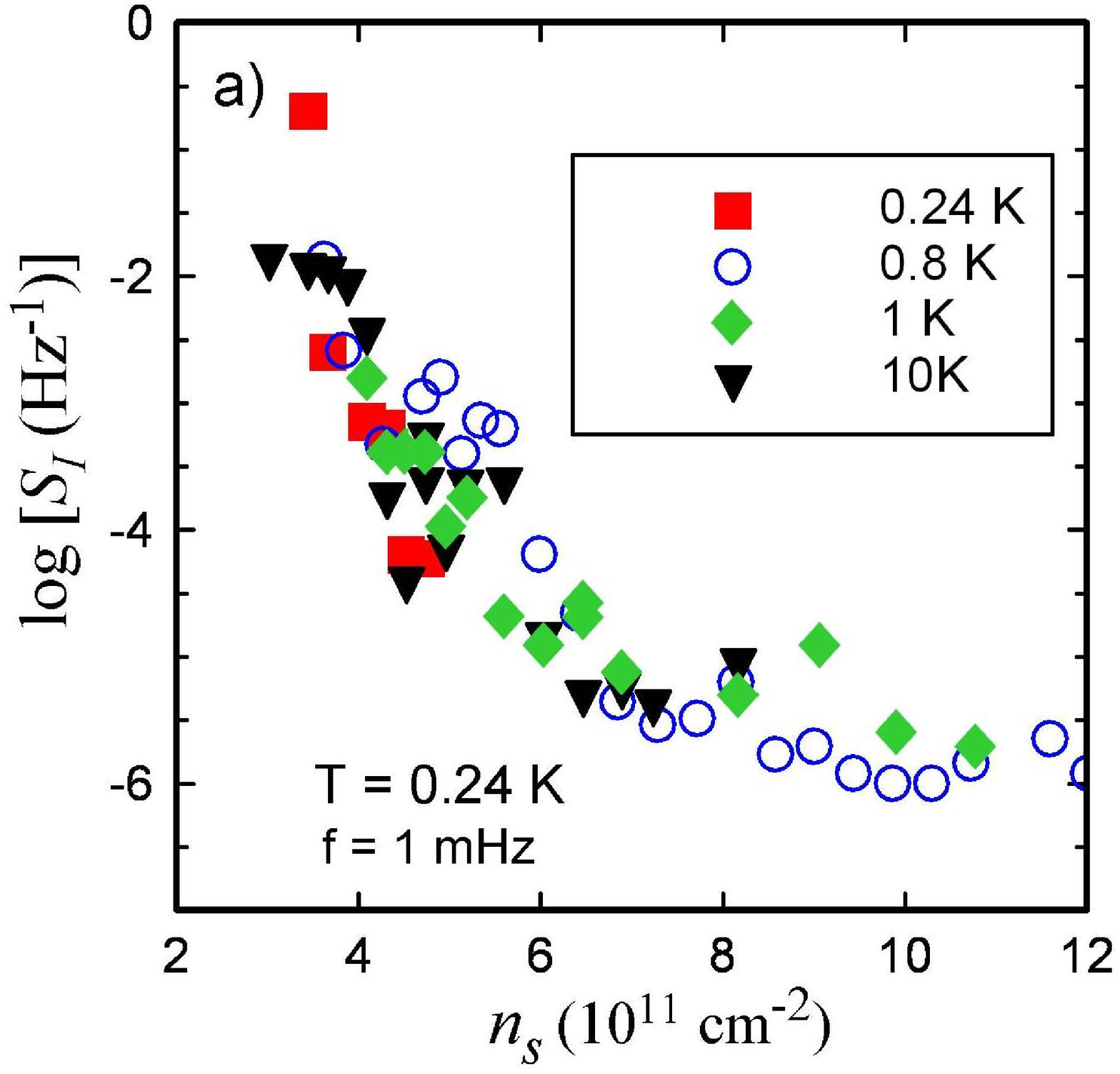}
\includegraphics[width=7.1cm]{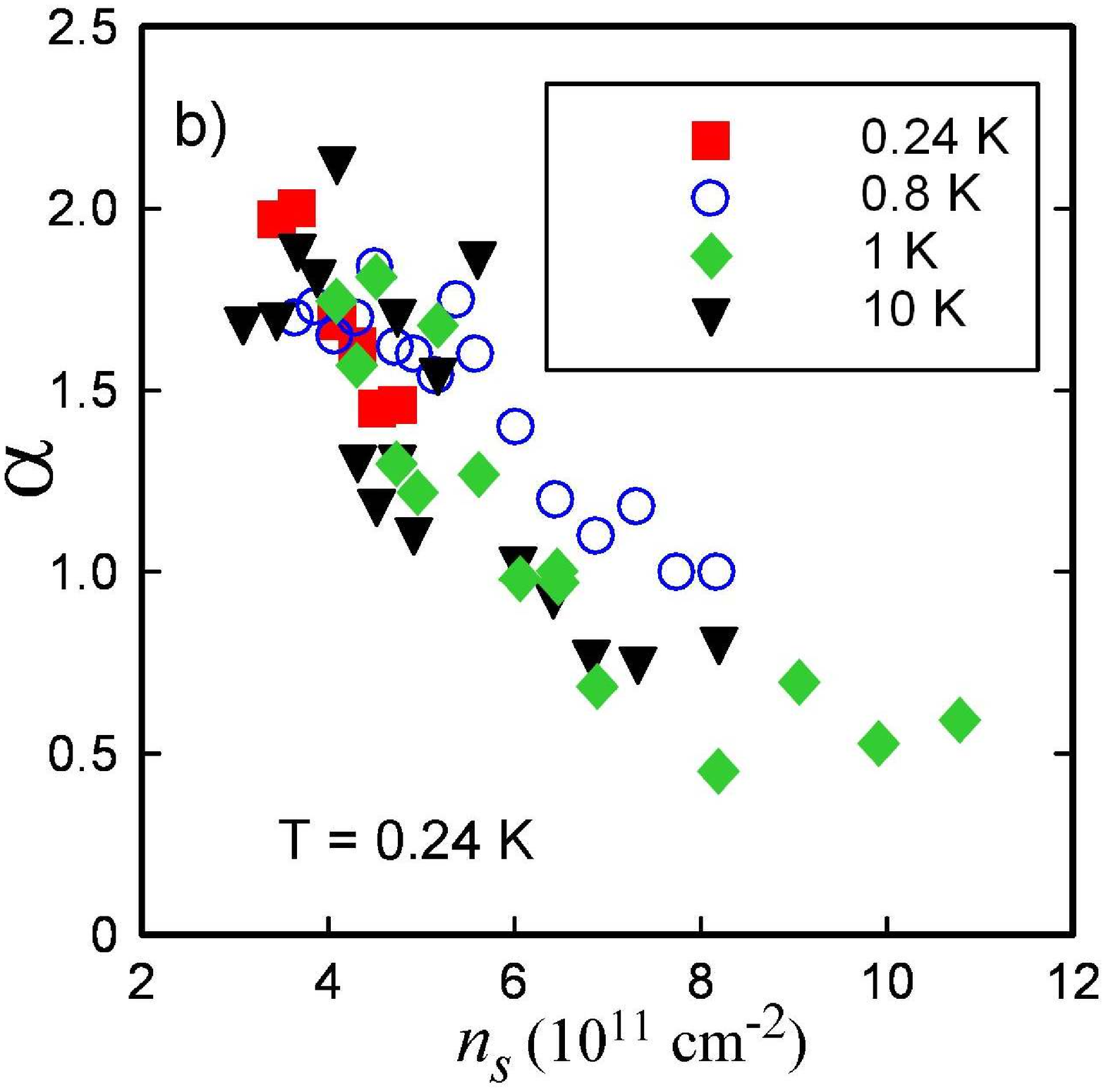}
\caption {(Color online)  Open symbols: sample A1; solid symbols: sample B.  Different symbols correspond to different $T$, as shown, where $n_s$ was changed before cooling down to 0.24~K to measure noise.  (a) The normalized noise power $S_{I}(f=1$~mHz) vs $n_s$.  (b) The exponent $\alpha$ vs $n_s$.}\label{fig:Salpha}
\end{figure}
%

The normalized power spectra of the relative changes in conductivity $\Delta\sigma(t)/\langle\sigma\rangle$ are found to obey the empirical law $S_I \propto 1/f^{\alpha}$ ($f$ -- frequency).  In order to compare the noise magnitudes under different conditions, the spectra were averaged over octaves, and the resulting fraction of power $S_I(f=1~\mathrm {mHz})$ is taken as the measure of noise.  Figure~\ref{fig:Salpha}(a) shows that $S_I(f=1~\mathrm {mHz})$ increases by several orders of magnitude as $n_s$ is reduced below $\sim 7\times 10^{11}$~cm$^{-2}$, while $\alpha$ rises rapidly to a much higher value [Fig.~\ref{fig:Salpha}(b)] in the same range of $n_s$.  Both effects indicate that the electron dynamics suddenly and dramatically slows down for $n_s<n_g\sim 7\times 10^{11}$~cm$^{-2}$, which is precisely the regime where non-Gaussian behavior of the noise is also observed (e.g. Fig.~\ref{fig:fig2}).  Therefore, the results of the noise measurements obtained using a $\Delta T$ protocol (black triangles in Fig.~\ref{fig:Salpha}) also provide evidence for glassy freezing of the 2DES for $n_s<n_g$, with the value of $n_g$ in agreement with that obtained from $\Delta n_s$ protocols.

\begin{figure}
\includegraphics[width=7.1cm]{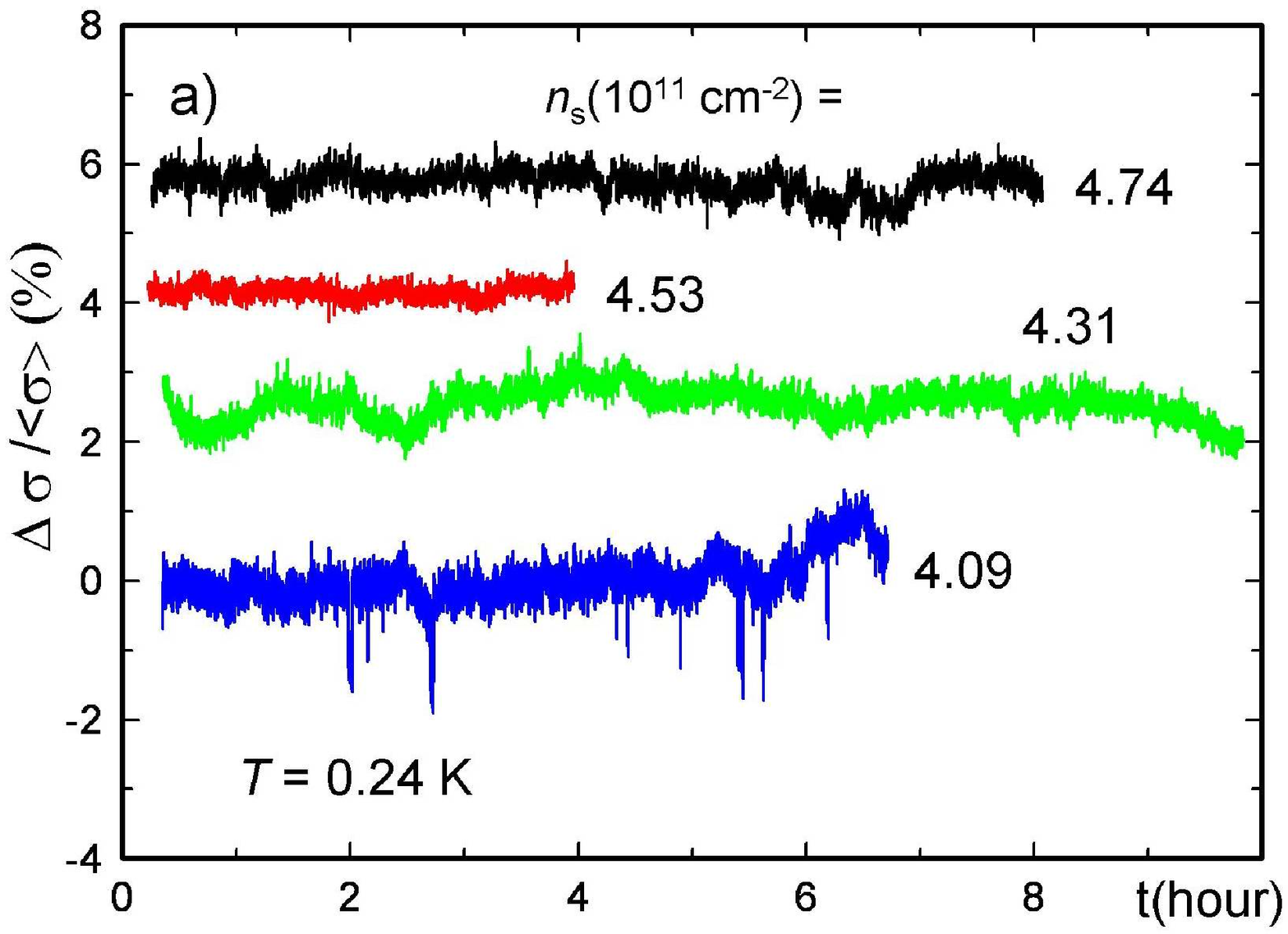}
\includegraphics[width=7.1cm]{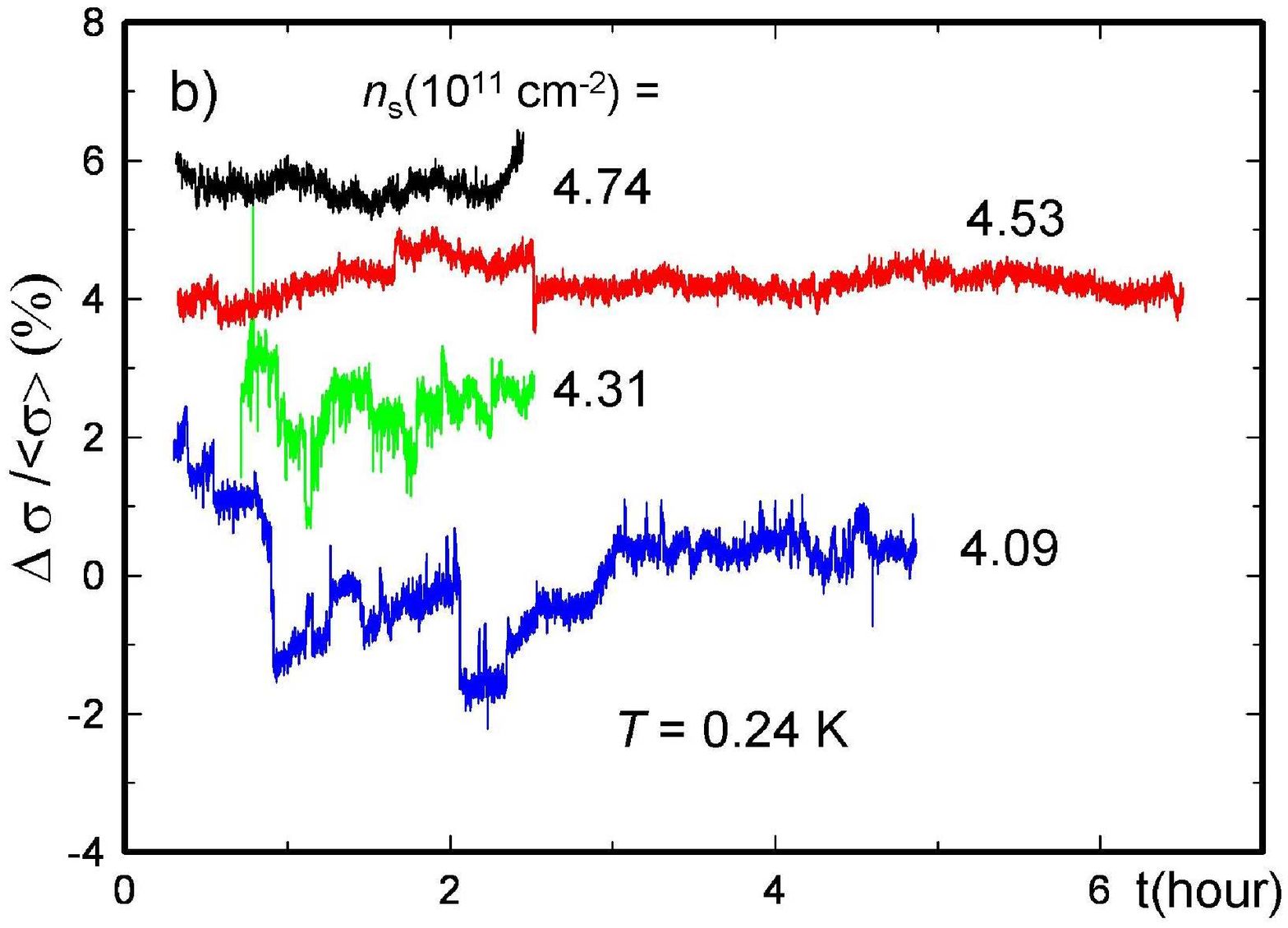}
\includegraphics[width=7.1cm]{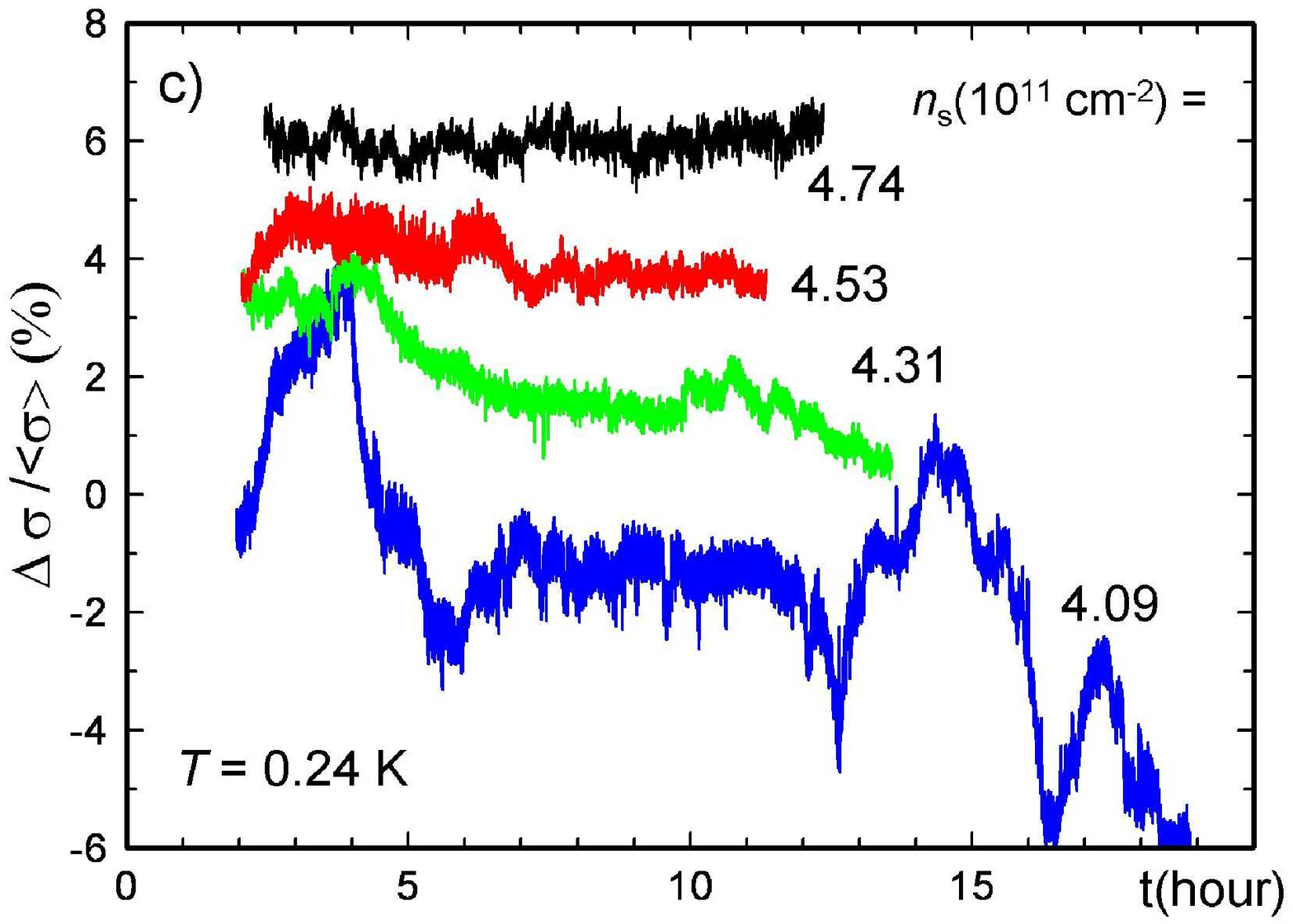}
\caption {(Color online) Sample B.  Conductivity $\sigma$ vs time at $T=0.24$~K for several $n_s$.  Traces are shifted for clarity.  The data were obtained (a) after changing $n_s$ at 10~K, followed by a slow decrease of $T$
for about 12 hours ($\Delta T$ protocol), and (b) after a small $n_s$ change ($\Delta n_s\approx 0.22\times 10^{11}$cm$^{-2}$) at low $T=0.24$~K ($\Delta n_s$ protocol).  In another run, (c) $n_s$ was changed at 1~K ($\Delta n_s$ protocol); after staying at 1~K for two hours, $T$ was lowered to 0.24~K in one hour, and $\sigma(t)$ was then measured.}\label{fig:protocols}
\end{figure}

Furthermore, the values of $S_I(f=1~\mathrm {mHz},n_s)$ and $\alpha(n_s)$ have been compared to
those from a $\Delta n_s$ protocol (see, e.g., Fig.~\ref{fig:protocols}).  In the latter, $n_s$ was changed
%
%
in small steps (typically $\sim 0.2\times 10^{11}$~cm$^{-2}$) at low $T$, where the equilibration times exceed the experimental time scales.\cite{Jan2006}  Figure~\ref{fig:Salpha} includes the results for $S_I$ and $\alpha$ obtained by changing $n_s$ at 1~K, 0.8~K (as in Ref.~\onlinecite{Snezana2002}) and 0.24~K.  Both the noise magnitude and the exponent $\alpha$ demonstrate that $\Delta T$ and \emph{small} $\Delta n_s$ protocols produce the same results.

While both protocols, cooling with a fixed $n_s$ and a small change of $n_s$ at low $T$, provide evidence for the onset of glassy dynamics from the fluctuations of $\sigma(t)$, we note that there are no visible relaxations of $\sigma$ following either process.  In both cases, though, the size of the perturbation ($k_{B}\Delta T\sim 10$~K or $\Delta E_{F}\sim 1-2$~K, respectively, where $E_F$ is the Fermi energy) is much smaller than $E_F$ ($\sim 20-60$~K),\cite{Ef-comment} i.e. probably too small to lead to any observable relaxations.  The relaxations were indeed observed for much larger changes of $n_s$,\cite{Jan2006,Jan2007-1,Jan2007-2} such that $\Delta E_{F}\sim 30-150$~K, i.e. for $\Delta E_{F}> E_F\gg k_{B}T$.

Since $\Delta E_{F}>k_{B}T$ in all $\Delta n_s$ protocols on a 2DES in Si MOSFETs, the perturbative treatments of the effect of the gate voltage,\cite{Markus2005-1,Markus2005-2} which may be relevant for other materials, are not suitable for this system.  In particular, here the applied $\Delta V_g$ (i.e. $\Delta E_{F}$) are expected to trigger major, \emph{collective} rearrangements of the electron configuration.  As shown below (Sec.~\ref{pdf-small}), a comparison of the full distribution of the fluctuations obtained from the two protocols demonstrates that even a small $\Delta n_s$ ($\Delta E_F\ll E_F$) has a qualitatively different effect than $\Delta T$, even though the results for the noise power spectra (Fig.~\ref{fig:Salpha}) and $\langle\sigma(n_s,T)\rangle$ are the same within experimental error.

\section{Distribution functions of the fluctuations}
\label{pdf}

In systems that are in equilibrium and away from criticality, the probability density functions (PDFs) of fluctuations in global quantities are Gaussian, according to the central limit theorem, which is based on the hypothesis that the system may be decomposed into many uncorrelated elements.  When this condition is not satisfied, e.g. due to the divergence of the correlation length at criticality or slow, nonexponential decay of correlation functions in glassy systems, non-Gaussian PDFs have been observed in a number of disparate systems.

\subsection{PDFs from $\Delta n_s$ and $\Delta T$ protocols}
\label{pdf-small}

\begin{figure}
\includegraphics[width=8.1cm]{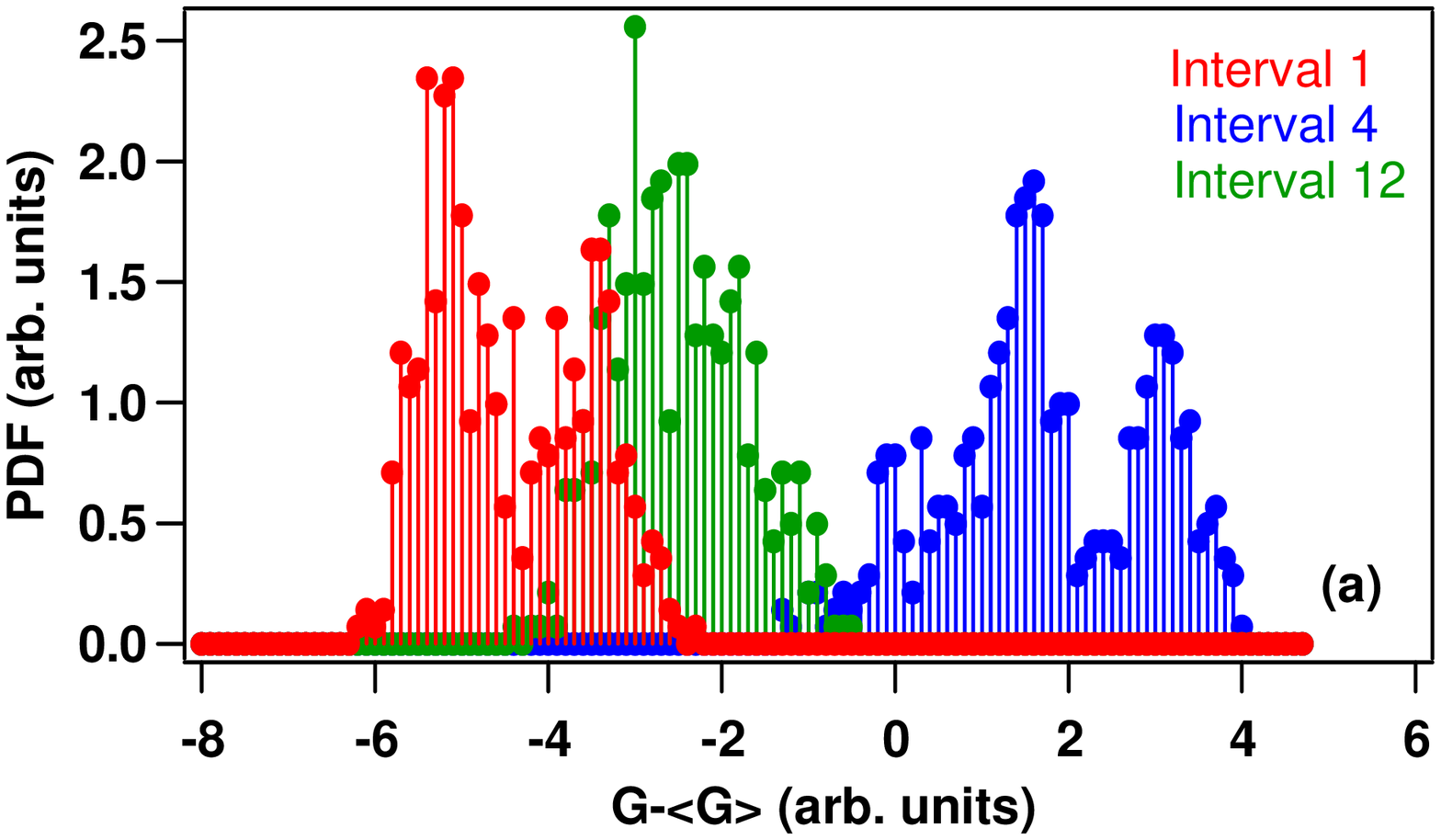}
\includegraphics[width=8.1cm]{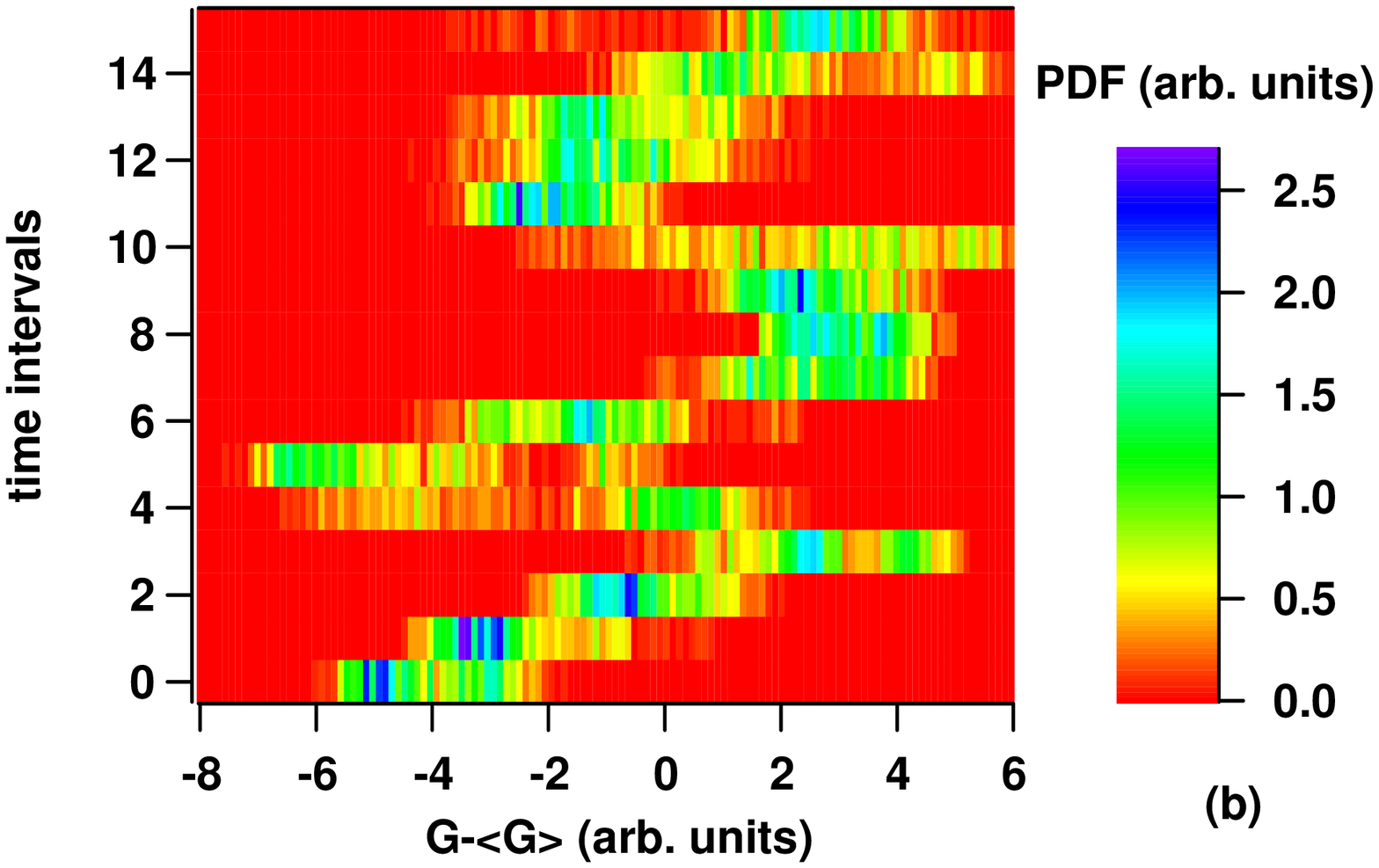}
\caption {(Color online) Sample A1.  Probability density function (PDF) of the conductance ($G$) fluctuations for $n_c<n_{s}(10^{11}$cm$^{-2})=5.58<n_g$ at $T=0.13$~K.  $n_s$ was fixed at $T\approx 0.8$~K after a small change $\Delta n_s=0.43\times 10^{11}$cm$^{-2}$, corresponding to $\Delta E_{F}\sim 3$~K$\ll E_{F}\approx 41$~K.  The measurement time was divided into sixteen intervals of about 35 minutes each.  PDFs for intervals 1, 4, and 12 are shown in (a), demonstrating the non-Gaussian behavior of the noise. (b)  The color map of the PDF of the fluctuations in $G$ as a function of all sixteen sequential time intervals.  The averaging $\langle\ldots\rangle$ was performed over the total measurement time ($\sim 9$~hours).}\label{fig:pdf}
\end{figure}

Figure~\ref{fig:pdf} shows some typical PDFs of the conductance ($G$) fluctuations for $n_s<n_g$ measured in a $\Delta n_s$ protocol.  The density $n_s$, lying in the metallic glass regime ($n_c<n_{s}<n_g$), was fixed at
%
%
$T\approx 0.8$~K after a small change $\Delta n_s=0.43\times 10^{11}$cm$^{-2}$ ($\Delta E_{F}\sim 3$~K$\ll E_{F}\approx 41$~K).  The sample was then cooled down to $T=0.13$~K, and $G(t)$ was recorded over sixteen sequential intervals of about 35 minutes each.  The PDFs, shown in Fig.~\ref{fig:pdf}(a) for three out of sixteen intervals, are clearly not Gaussian, but rather have a complicated, multi-peaked structure that changes with time.

These results indicate that, on experimental time scales, the 2DES ``visits'' only a small number of states, but that new states also become available with time.  This is also reflected in the correlated ``wandering'' of the PDFs with time [Fig.~\ref{fig:pdf}(b)].  For $n_s>n_g$, on the other hand, the PDFs become Gaussian (not shown) on much shorter time scales.

The PDFs obtained from a $\Delta T$ protocol (Fig.~\ref{fig:pdfT}, left column) exhibit the same general behavior:
%
\begin{figure}
\includegraphics[width=8.1cm]{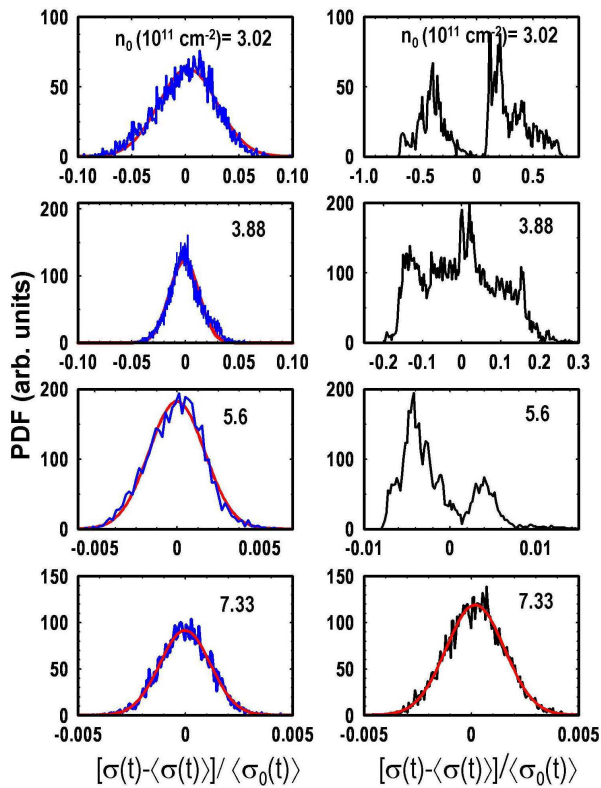}
\caption {(Color online)  Sample A.
PDF vs $[\sigma(t)-\langle\sigma(t)\rangle]/\langle\sigma_0\rangle$ at $T=0.24$~K for several $n_0$, as shown.  $\langle\sigma_0\rangle$ is the time-averaged conductivity corresponding to $n_0$ at the measurement $T$, obtained from a $\Delta T$ protocol.  The red curves are fits to a Gaussian distribution.  Left column: PDFs of the noise after cooling from 10~K to 0.24~K for each given $n_0$ ($\Delta T$ protocol).  The cooling time was at least one hour long. In this protocol, $\langle\sigma(t)\rangle\equiv\langle\sigma_0\rangle$, i.e. there are no observable relaxations after cooling.  Right column: PDFs of the noise measured after a subsequent change of $n_s$ from $n_0$ to a much higher value $n_1=20.26\times 10^{11}$cm$^{-2}$ during $t_w=1000$~s.  In this protocol, $\langle\sigma(t)\rangle$ describes the slowly relaxing background.}\label {fig:pdfT}
\end{figure}
%
they are non-Gaussian for $n_s<n_g$ and Gaussian for $n_s>n_g$.  (Even though the non-Gaussian shape of the PDFs for $n_s<n_g$ is more obvious on a semi-log scale, they are presented on a linear scale for easier comparison to the effects of $\Delta n_s$.  See also Sec.~\ref{exc}.) Here, however, even the non-Gaussian PDFs are smooth, single-peaked functions, reminiscent of PDFs in a variety of systems displaying critical,\cite{criticalPDF1,criticalPDF2} glassy,\cite{hetero-review} or other out-of-equilibrium behavior (e.g. the Danube water level\cite{Danube}).  In all these systems, the PDFs are skewed, resembling a zero-centered Gaussian, which describes pseudo-equilibrium fluctuations, with one (exponential) tail that is due to large, rare events.
In that context, the totally irregular structure of the PDFs obtained after a small $\Delta n_s$ (Fig.~\ref{fig:pdf}) is striking.

The above differences can be understood in the following way.  After cooling, the 2D charge glass settles in a deep energy valley, perhaps a global minimum of the FEL, corresponding to a given $n_s$.  As a function of time, the system will explore the subvalleys of the energy landscape, but it will in general remain in the ``main'', deep valley.  Such pseudo-equilibrium fluctuations will produce PDFs of the type shown in the left column of Fig.~\ref{fig:pdfT}.  In the $\Delta n_s$ protocol, in contrast, a change in the number of charge carriers reshuffles all energies because of the Coulomb interactions, so that the 2DES in general no longer finds itself in the minimum of the FEL.  This will result in a more dramatic wandering of the system through the FEL, producing the PDFs of the type shown in Fig.~\ref{fig:pdf}.

\subsection{PDFs from a waiting-time protocol}
\label{pdf-large}

Large $\Delta n_s$ ($\Delta E_F > E_F$) lead to observable relaxations in $\sigma (t)$.  This effect has been used to investigate aging and memory in a 2DES in Si by employing the so-called waiting-time ($t_w$) protocol,\cite{Jan2007-1,Jan2007-2,Jan2009} which consists of the following.  It starts with a $\Delta T$ protocol, i.e. cooling at a fixed density $n_0$ and measuring noise (see left column in Fig.~\ref{fig:pdfT} for the corresponding PDFs).  The density is then switched rapidly (within 1 s) to a different value $n_1$, where it is kept for a time $t_w$.  During that time, the system relaxes away from its initial (pseudo-)equilibrium state determined by $n_0$ and towards a new equilibrium state determined by $n_1$.  Finally, the density is changed back (within 1 s) to $n_0$ at $t=0$, and the slowly evolving $\sigma(t)$ is measured.  We note that here $t=0$ is defined as the time when the charge carrier density reattains its original value $n_0$.  If $t_w$ is shorter than the equilibration time, then $\sigma (t)$ will depend on $t_w$, i.e. on the measurement history (aging efect).  It is also said that the system has a memory of the time it spent with $n_1$.  Here we focus on that situation, i.e. on noise during aging.

Figure~\ref{fig:pdfT} presents a comparison of the PDFs obtained from $\Delta T$ (left column) and $t_w$ (right column) protocols for different $n_0$ and the same $n_1=20.26\times 10^{11}$cm$^{-2}$ and $t_w=1000$~s.  For the highest $n_0\gtrsim n_g$, the PDFs are Gaussian in both protocols, with roughly the same standard deviation.  For lower $n_0$, however, there are some remarkable differences.  First, the noise after a temporary density change is much larger than the noise before the change, that is, after a $\Delta T$ protocol.  At the lowest $n_0$, this difference amounts to more than one order of magnitude, as first noted in Ref.~\onlinecite{Jan2007-2}.  Second, the change of density during $t_w$ results in complex, multi-peaked, totally random-looking PDFs, analogous to those obtained from $\Delta n_s$ protocols (Sec.~\ref{pdf-small}).  These differences are even more striking considering that, for a given $n_0$, the PDFs in Fig.~\ref{fig:pdfT} were obtained from the measurements that were carried out under \emph{exactly the same} experimental conditions; the only difference was in the sample history.

These results imply the existence of a rugged free energy landscape that is modified in a non-trivial way by all $\Delta n_s$ of practical interest (corresponding to $\Delta E_F >k_{B}T$).  The rearrangements of the electron configuration and the reshuffling of the FEL caused by Coulomb interactions will clearly need to be taken into account in theoretical models of the glassy dynamics of the 2DES in Si.  This complicated problem is further aggravated by the proximity to the MIT, where substantial changes in the screening of the 2DES may be expected.\cite{Pastor,Markus2005-PRL,Pankov2005}

\begin{figure}
\includegraphics[width=8.1cm]{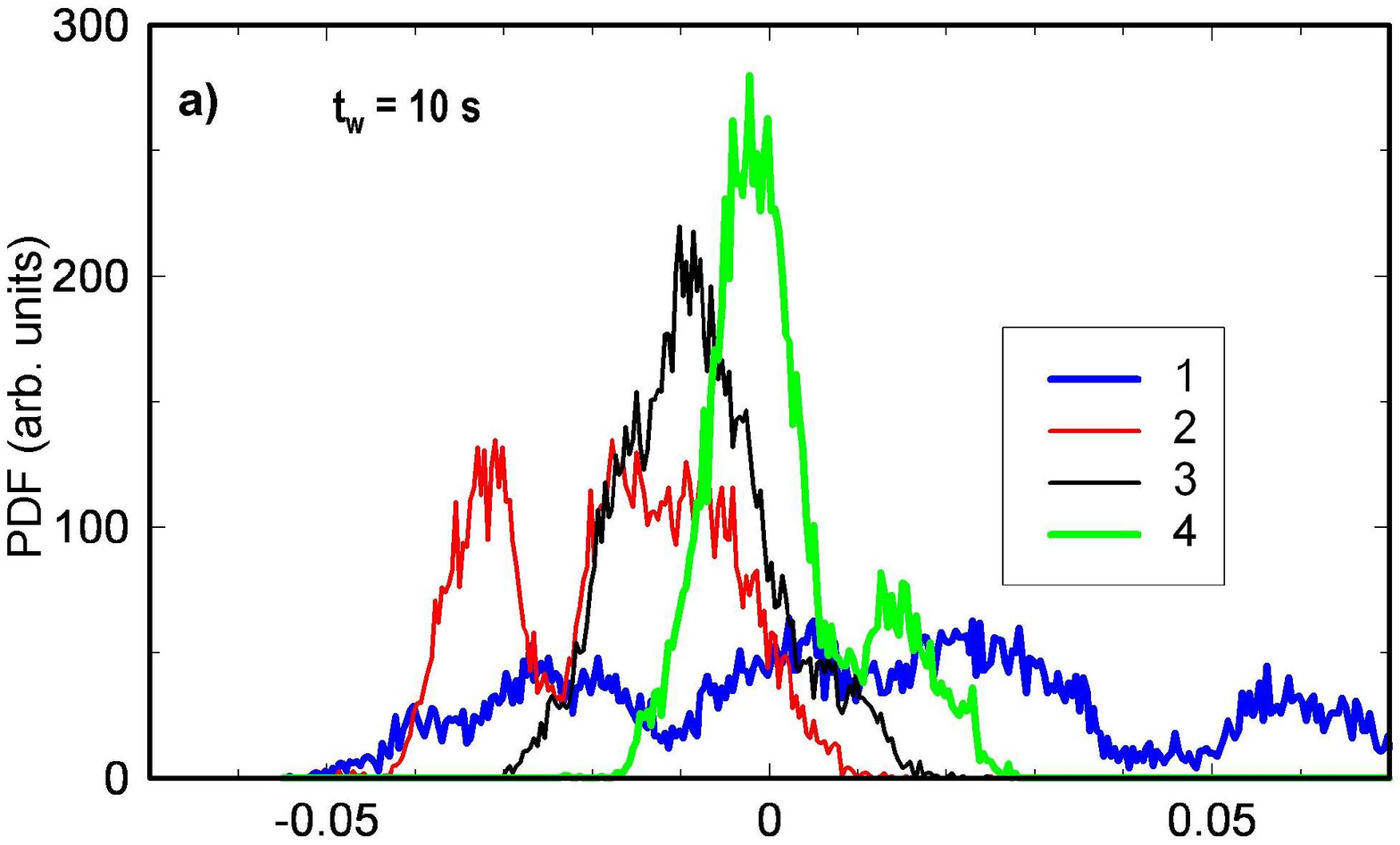}
\includegraphics[width=8.1cm]{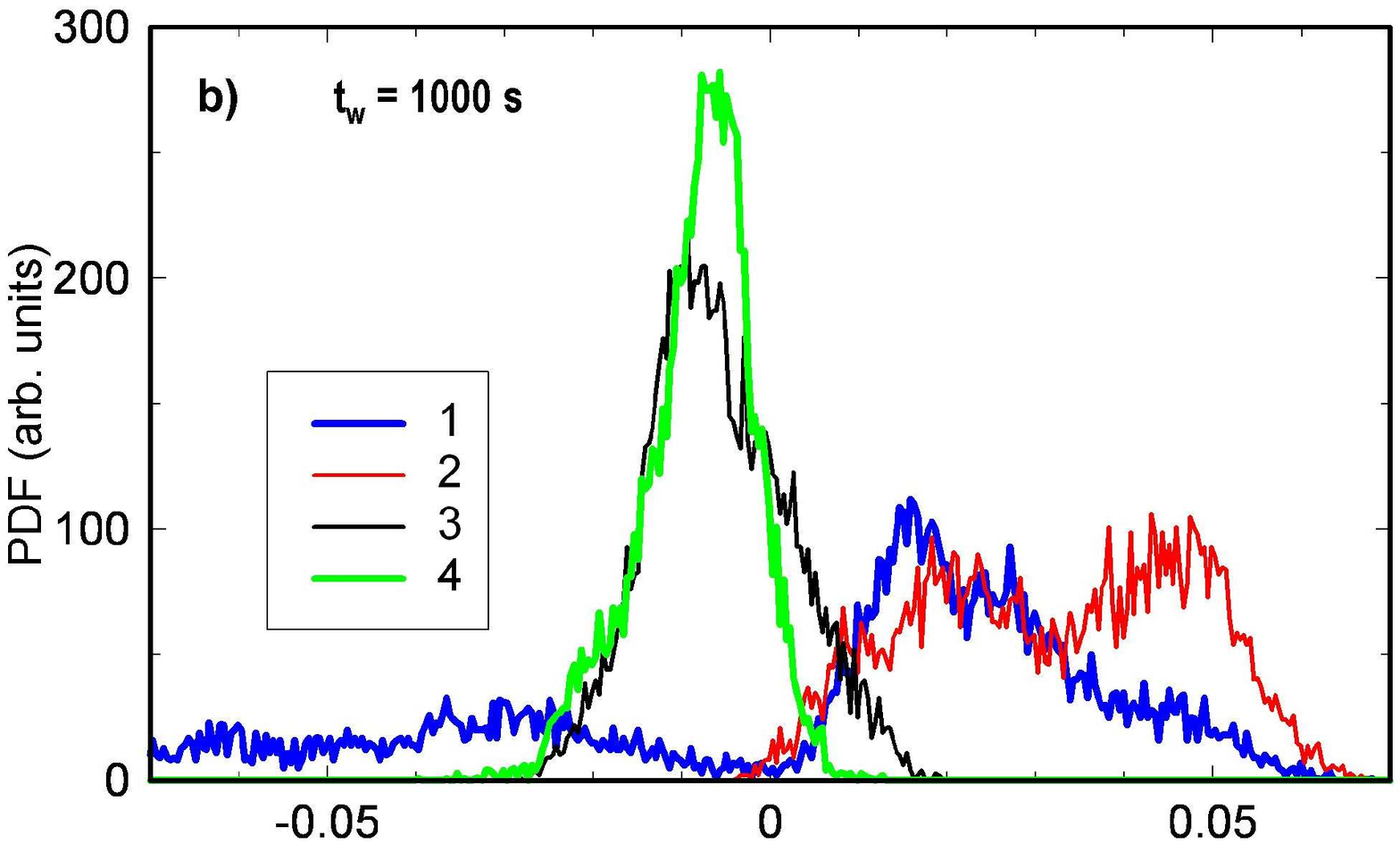}
\includegraphics[width=8.1cm]{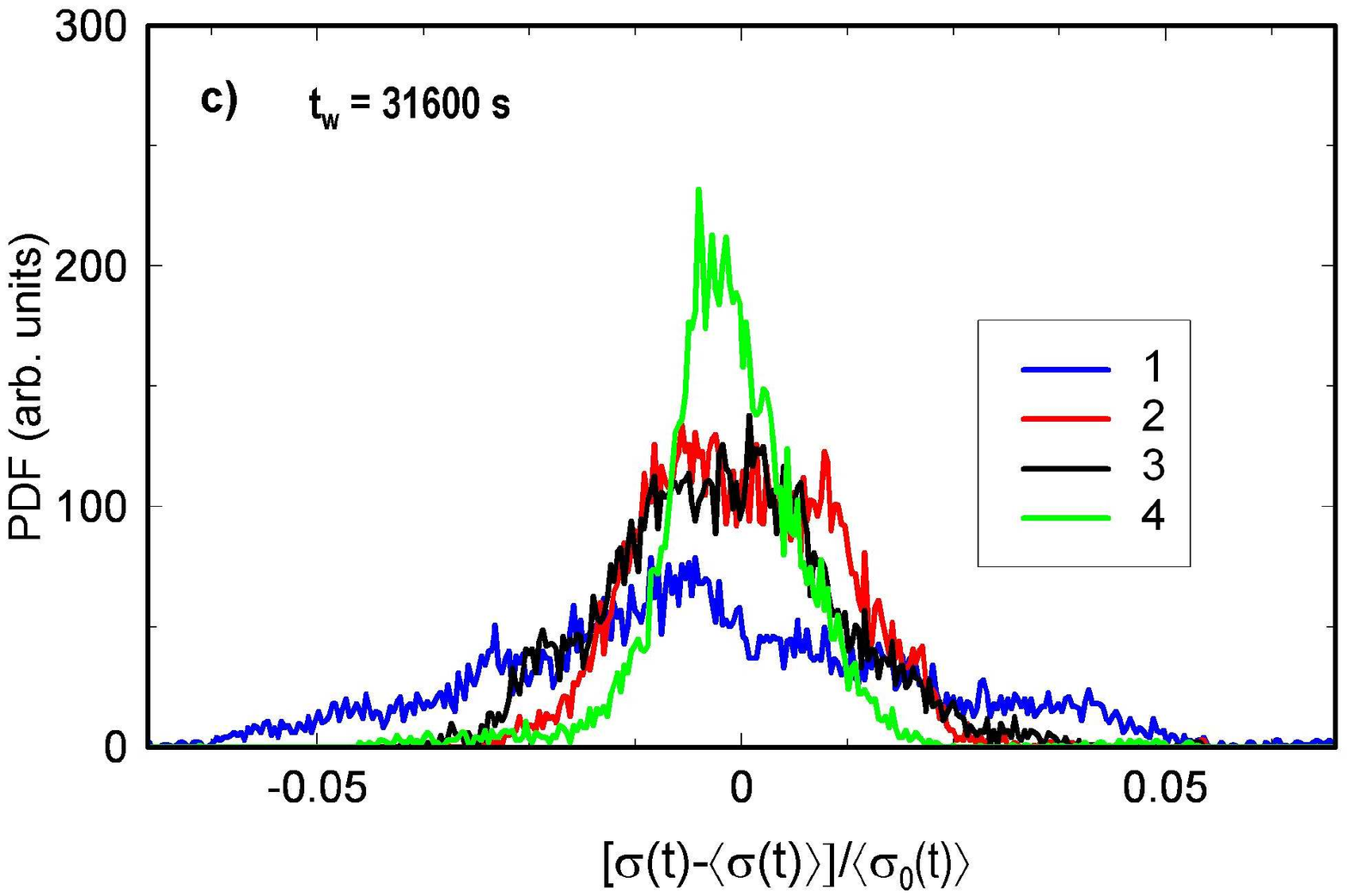}
\caption {(Color online)  Sample A.  PDF \textit{vs.} $[\sigma(t)-\langle\sigma(t)\rangle]/\langle\sigma_0\rangle$ at $T=1$~K for $n_0=3.02\times 10^{11}$cm$^{-2}$, following a change of $n_s$ to $n_1=20.26\times 10^{11}$cm$^{-2}$ during (a) $t_w=10$~s, (b) $t_w=1000$~s, and (c) $t_w=31600$~s. For each $t_w$, $n_0$ was initially set at 10~K.  Different curves correspond to PDFs for four consecutive 3-hour time intervals, as shown.  The PDFs become more Gaussian with time, as the system ages.}\label{fig:pdftw}
\end{figure}

The large noise that accompanies the aging process reflects the collective wandering of the 2DES through the FEL, as the system relaxes towards a global minimum that corresponds to the equilibrium state.  In other out-of-equilibrium systems (e.g. polymers and colloidal glasses,\cite{Ciliberto-PDF}, conventional spin glasses\cite{SG-PDF}), the PDFs have been observed to evolve towards a Gaussian shape as the system ages.  Here the evolution of the PDFs has been investigated from the data obtained in several consecutive time intervals, during the relaxation of $\sigma(t)$ induced using a $t_w$ protocol, as described above.  A typical behavior is presented in Fig.~\ref{fig:pdftw}, which
%
%
shows the PDFs for the same density $n_0=3.02\times 10^{11}$cm$^{-2}$ in the insulating regime ($n_0<n_c$) and for three different $t_w$.  In each case, the PDFs for four consecutive 3-hour time intervals demonstrate that the noise indeed becomes more Gaussian with time, as the system slowly approaches equilibrium.

Understanding of the age dependence of the PDFs
is a non-trivial problem in general.  Even in situations where PDFs feature a ``simple'' zero-centered Gaussian with an exponential tail, different models have been proposed.  In addition, numerical simulations of aging models have been performed, including those based on energy landscapes with hierarchically organized metastable states (see, e.g., Ref. \onlinecite{Sibani-hier}).  While it is unclear whether any of that work is relevant to our system, we note that resistance noise measurements on both spin glasses\cite{Weissman93} and the Coulomb glass in a 2DES in Si\cite{Jan2002} have provided support for the hierarchical picture of glassiness.\cite{Binder}  A number of other, remarkable similarities in the phenomenological behavior of these two types of glassy systems have been also reported.\cite{Jan2006,Jan2007-1,Jan2007-2}

\section{Dependence of conductance noise on applied electric field}
\label{exc}

The emergence of the non-Gaussian conductance noise for $n_s<n_g$, i.e. in precisely the same regime where relaxations exhibit the out-of-equilibrium behavior,\cite{Jan2006,Jan2007-1,Jan2007-2,Jan2009} as well as the detailed study and comparison of the noise obtained using different protocols (Secs.~\ref{noisepower} and \ref{pdf}), clearly demonstrate that it reflects the intrinsic glassiness of the 2DES.  For example, the qualitatively different behavior of the PDFs measured under the same conditions, but with different sample histories (Fig.~\ref{fig:pdfT}), rules out any spurious effects as possible sources of non-Gaussian noise from a $t_w$ protocol.  Likewise, the agreement of the results for the noise power spectra obtained using $\Delta T$ and $\Delta n_s$ protocols (Fig.~\ref{fig:Salpha}) implies that they have the same origin.  Experiments on a disordered indium oxide, an electron glass deep in the insulating regime, have shown that,\cite{Zvi-nonlinear} in that system, the degree of non-Gaussian behavior may be affected by the applied electric bias $V_{\mathrm{exc}}$.  In particular, the non-Gaussian nature of the noise was found to be non-monotonic with $V_{\mathrm{exc}}$, vanishing at both low and high bias, and peaking at an intermediate value of $V_{\mathrm{exc}}$.  This behavior was attributed to a nonlinear effect inherent to variable-range hopping transport.  While the indium oxide samples were indeed very deep in the hopping regime, this is in general not true of the experiments on a 2DES, where glassiness sets in already on the metallic side of the MIT ($n_g>n_c$).  Nevertheless, it is interesting to investigate whether similar effects might be observed also in the glassy regime of the 2DES, in particular in the range of parameters where non-Gaussian conductance noise has been studied (Secs.~\ref{noisepower}, \ref{pdf} and Refs.~\onlinecite{Snezana2002,Jan2002,Jan2004}).

\begin{figure}
\includegraphics[width=8.1cm]{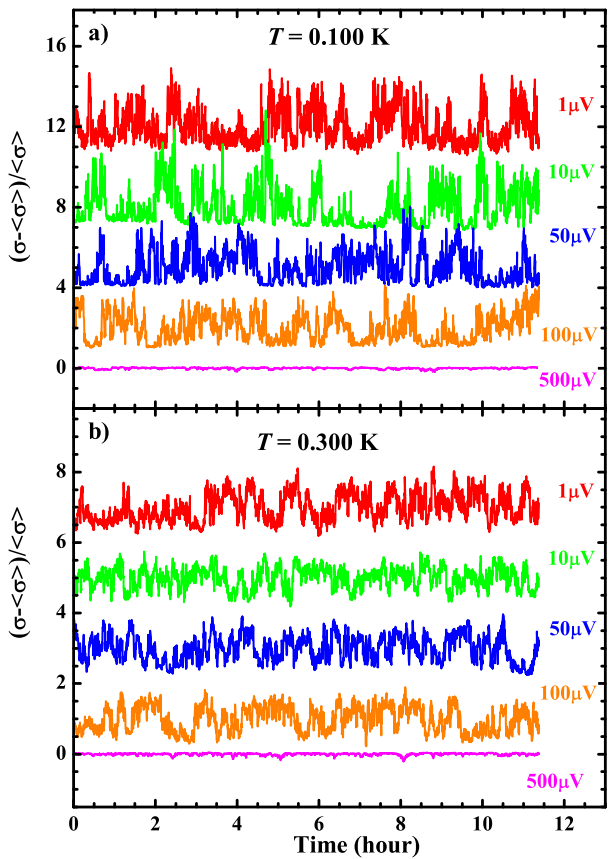}
\caption {(Color online) Sample B1. Relative fluctuations of conductivity $(\sigma(t)-\langle\sigma\rangle)/\langle\sigma\rangle$ vs time for $n_s=3.24 \times 10^{11}$~cm$^{-2}<n_c$ and different $V_{\mathrm{exc}}$ at a) $T=0.100$~K and b) $T=0.300$~K.  Traces are shifted for clarity. The data were obtained after changing $n_s$ at $T \approx 20$~K, cooling down to the desired $T$, and then measuring $\sigma(t)$ for different values of $V_{\mathrm{exc}}$.}\label{fig:rawVexc}
\end{figure}

In order to explore the effects of $V_{\mathrm{exc}}$, conductance noise $\Delta\sigma(t)/\langle\sigma\rangle$ was measured using a $\Delta T$ protocol.  After changing $n_s$ at a high $T$ (here at $\approx 20$~K), the 2DES was cooled to a desired $T$, and $\sigma (t)$ measured for constant values of $V_{\mathrm{exc}}$ ranging from 0.5~$\mu$V to 1~mV.  Typical data are presented in Fig.~\ref{fig:rawVexc}, which shows $\Delta\sigma/\langle\sigma\rangle$ vs
%
%
time for $n_s=3.24 \times 10^{11}$~cm$^{-2}$ in the insulating regime ($n_s<n_c$) and for several $V_{\mathrm{exc}}$ at two different $T$.  The noise is obviously non-Gaussian and large, of the order of 100$\%$ at low $T$.  The amplitude of the relative fluctuations decreases with increasing $T$, as established in earlier studies,\cite{Snezana2002,Jan2002} as well as for high enough excitations.

\begin{figure}
\includegraphics[width=8.1cm]{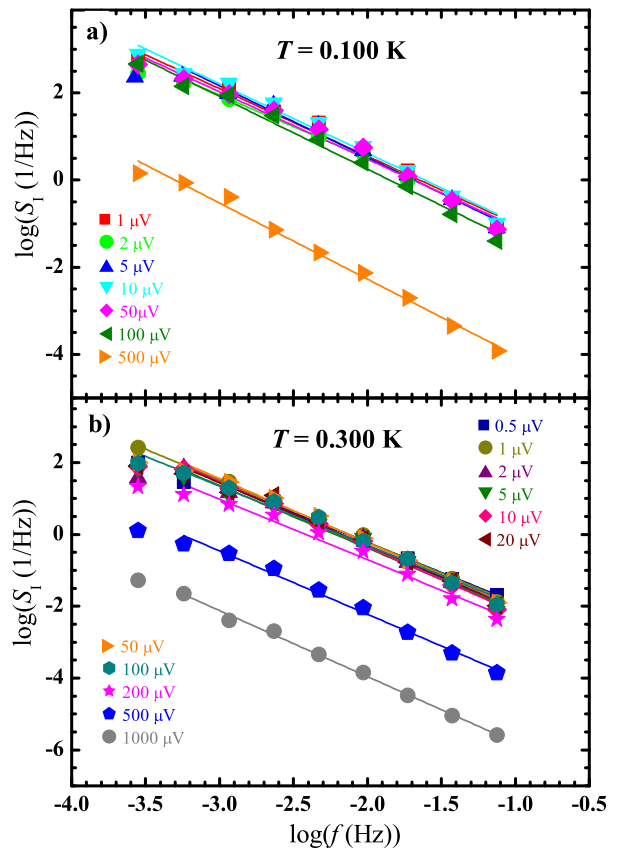}
\caption {(Color online) Sample B1. The octave-averaged noise power spectra $S_{I}=C/f^{\alpha}$ vs $f$ for several $V_{\mathrm{exc}}$ at a) $T=0.100$~K and b) $T=0.300$~K.  Solid lines are linear least-squares fits with the slopes equal to $\alpha$.}\label{fig:powerexc}
\end{figure}

\begin{figure}
\includegraphics[width=8.1cm]{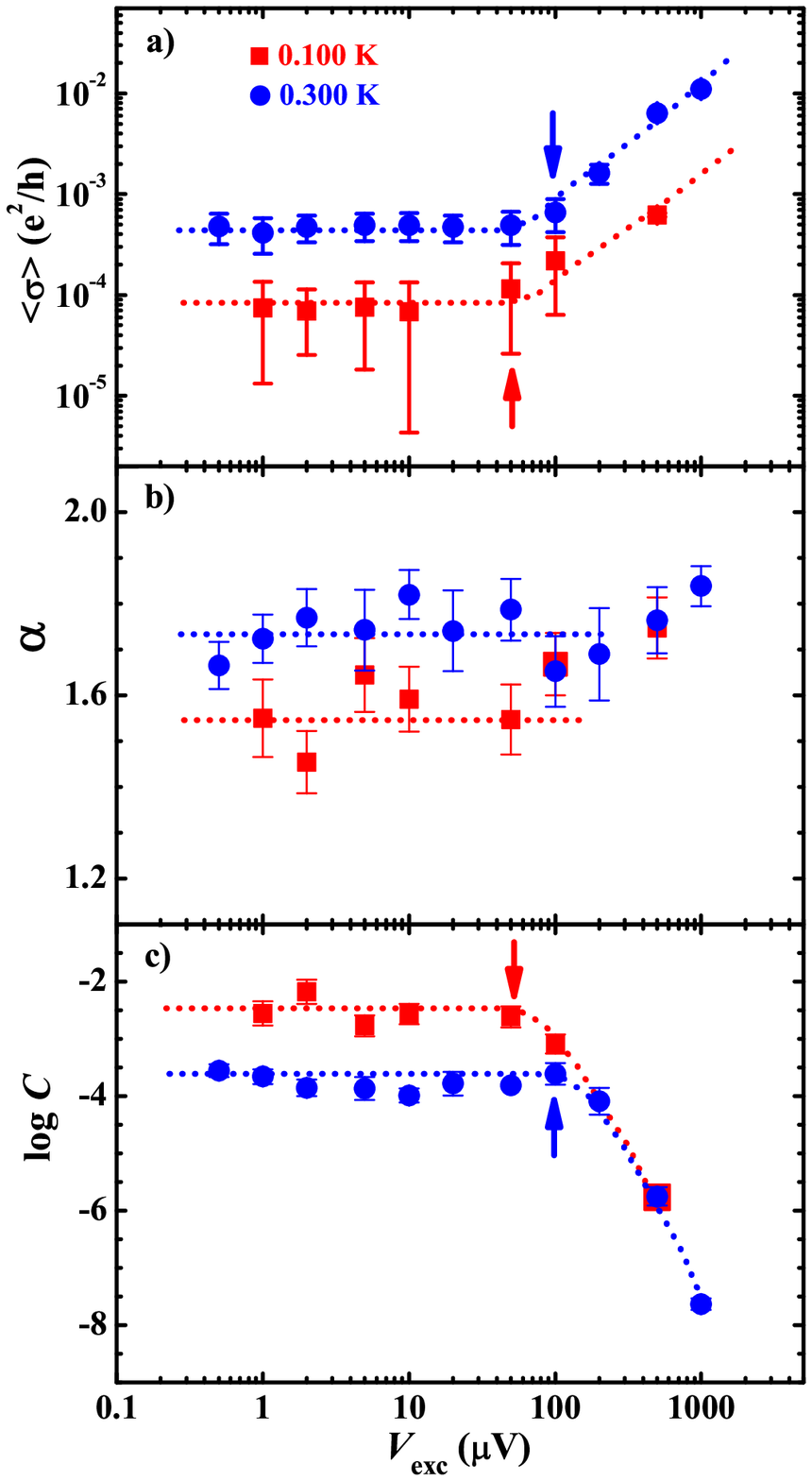}
\caption {(Color online) Sample B1.  (a) $\langle\sigma\rangle$, (b) exponent $\alpha$, and (c) parameter $C$ (in $S_I=C/f^{\alpha}$) vs $V_{\mathrm{exc}}$ at $T=0.100$~K (red squares) and $T=0.300$~K (blue dots) for $n_s=3.24 \times 10^{11}$~cm$^{-2}<n_c$.  In (a) and (c), the arrows show the onset of the non-Ohmic regime, and dotted lines guide the eye.  In (b), dotted lines represent the average values of $\alpha$ in the Ohmic regime: $\alpha=1.5\pm 0.2$ at $T=0.100$~K and $\alpha=1.7\pm 0.1$ at $T=0.300$~K.}\label{fig:sigmaexc}
\end{figure}

It is indeed obvious from the corresponding normalized power spectra $S_I(f)=C/f^{\alpha}$ (Fig.~\ref{fig:powerexc})
%
%
that the noise does not depend on the applied excitation at low enough $V_{\mathrm{exc}}$, but the power spectra $S_I$ become suppressed as $V_{\mathrm{exc}}$ is increased beyond a certain value.  A detailed comparison of the dependence of the exponent $\alpha$, the parameter $C$, and the time-averaged conductivity $\langle\sigma\rangle$ on $V_{\mathrm{exc}}$ is presented in Fig.~\ref{fig:sigmaexc}.  It is apparent that $\langle \sigma \rangle$ remains
%
%
Ohmic over a wide range of $V_{\mathrm{exc}}$ [Fig.~\ref{fig:sigmaexc}(a)], up to an electric field $\sim 50$~V/m ($V_{\mathrm{exc}}\sim 100~\mu$V) at $T\sim 0.1$~K (where $eV_{\mathrm{exc}}\sim k_{B}T$).  In the same range of $V_{\mathrm{exc}}$, both the exponent $\alpha\approx 1.6$ [Fig.~\ref{fig:sigmaexc}(b)] and the noise magnitude, i.e. the parameter $C$ [Fig.~\ref{fig:sigmaexc}(c)], also do not depend on $V_{\mathrm{exc}}$ within experimental error.  The latter result is expected if the applied voltage is sufficiently small to be in the Ohmic regime and if the noise is in fact due to conductance (or resistance) fluctuations.  At higher $V_{\mathrm{exc}}$, where $\langle\sigma\rangle$ deviates from the Ohmic behavior, $\alpha$ does not seem to change with $V_{\mathrm{exc}}$ or $T$ within the scatter of data, but the parameter $C$ clearly decreases rapidly with increasing $V_{\mathrm{exc}}$, similar to the effect of raising $T$.  Therefore, the results strongly suggest that the suppression of the noise by high $V_{\mathrm{exc}}$ is most likely caused by simple heating.

\begin{figure}
\includegraphics[width=8.1cm]{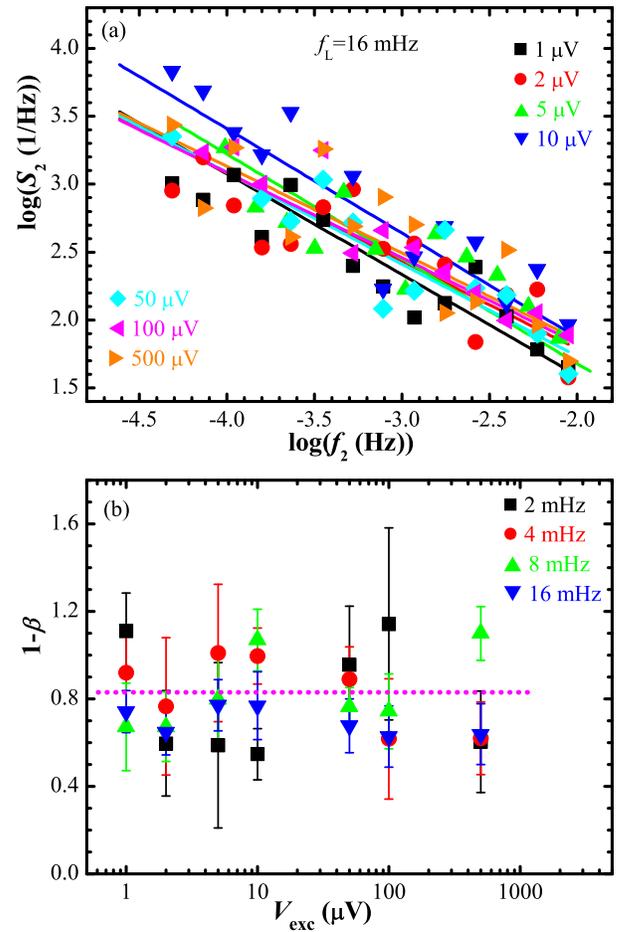}
\caption {(Color online) Sample B1, $T=0.100$~K, $n_s=3.24 \times 10^{11}$~cm$^{-2}$.  (a) Second spectral density $S_2 \propto 1/f_{2}^{1-\beta}$ vs $f_2$ for several values of $V_{\mathrm{exc}}$, as shown.  $S_2$ was measured in the frequency octave $f=(f_L,2f_L)$, where $f_L = 16$~mHz. The solid lines are linear least-squares fits with the slopes equal to $1-\beta$.  (b) Exponent ($1-\beta$) vs $V_{\mathrm{exc}}$ for different values of $f_L$, as shown. Dotted line guides the eye.}\label{fig:S2}
\end{figure}

Similar to previous studies of noise on the 2DES in Si\cite{Jan2002,Jan2004} and those in indium oxide\cite{Zvi-nonlinear}, the effect of $V_{\mathrm{exc}}$ on the non-Gaussian character of the noise has been investigated by analyzing the second spectrum $S_2(f_2,f)$, a fourth-order noise statistic.\cite{Weissman88,Weissman92,Weissman93}  In particular, $S_2(f_2,f)$ represents the power spectrum of the fluctuations of $S_{I}(f)$ with time.  For Gaussian noise, $S_2(f_2)$ is white (independent of $f_2$).  On the other hand, for non-Gaussian noise, $S_2 \propto 1/f_2^{1-\beta}$ with the exponent $(1-\beta)\neq 0$. Therefore, the value of the exponent $(1-\beta)$ is a convenient measure of the non-Gaussian character of the noise.  $S_2$ was analyzed by using digital filtering \cite{Seidler-Solin,Abkemeier} in a given frequency range $f=(f_L, f_H)$ (usually $f_H = 2f_L$).  Some examples of the normalized $S_2(f_2)$, with the Gaussian background subtracted, are presented in Fig.~\ref{fig:S2}(a) for the data in Fig.~\ref{fig:rawVexc}(a).  The second spectrum is clearly non-white, in
%
%
agreement with earlier work.\cite{Jan2002}  Here we find that the exponent $(1-\beta)\approx 0.8$ does not depend on $V_{\mathrm{exc}}$ [Fig.~\ref{fig:S2}(b)] in the regime studied.  Therefore, non-Gaussian noise is observed over a wide range of electric fields in the regime of linear or Ohmic response.

Detailed studies of non-Gaussian noise in the 2DES in Si (Secs.~~\ref{noisepower}, \ref{pdf} and Refs.~\onlinecite{Snezana2002,Jan2002,Jan2004}) have been performed with the applied electric fields $\sim 5$~V/m, i.e. an order of magnitude lower than the field where the non-Ohmic behavior sets in (see Fig.~\ref{fig:sigmaexc}).  This is in contrast to the experiment on indium oxide,\cite{Zvi-nonlinear}, where only the non-Ohmic regime was explored, and where the non-monotonic dependence of the non-Gaussian behavior on $V_{\mathrm{exc}}$ was reported.  A study of such a deep non-linear regime, however, is beyond the scope of this work, so the existence of a similar effect in our system for much higher $V_{\mathrm{exc}}$ cannot be ruled out.  Our study does demonstrate, on the other hand, that the non-Gaussian noise in a 2DES is not due to the same non-linear mechanism caused by the high bias.  Instead, it provides further evidence that the correlated noise reflects the intrinsic out-of-equilibrium dynamics of the 2DES in Si.

\section{CONCLUSIONS}
\label{concl}

We have presented a study of the conductance noise in a 2DES in Si at low $T$, measured by using three different experimental protocols.  In the so-called $\Delta T$ protocol, the carrier density $n_s$ is fixed at a high $T$ where the system is in an equilibrium state, and the noise is measured after cooling with a fixed $n_s$.  The results provide evidence that, similar to other types of glasses, the 2DES falls out of equilibrium as $T$ is reduced.  This glassy freezing is observed for all densities $n_s<n_g$, as $T\rightarrow 0$.

In a $\Delta n_s$ protocol, $n_s$ is changed by a small amount at such low $T$ that the system is unable to equilibrate on experimental time scales.  While the perturbations introduced by both $\Delta T$ and $\Delta n_s$ protocols are too small to lead to observable relaxations of $\sigma$ (i.e. $k_{B}\Delta T, \Delta E_F\ll E_F$), they produce the same results for the power spectra of the non-Gaussian noise observed for $n_s<n_g$, where the glass transition density $n_g$ is found to be the same in both cases.  The analysis of the probability density functions (PDFs) of the fluctuations, however, reveals that $\Delta n_s$ has a strikingly different effect on the system than $\Delta T$ (both $\Delta E_F, k_{B}\Delta T>k_{B}T$).  In particular, the results strongly suggest that the density change reshuffles all energies, because of the Coulomb interactions, thus modifying the free energy landscape of the 2DES.  For this reason, theoretical modeling of the glassy dynamics in this system might be considerably more difficult than in some other types of glassy materials.

The effect of the density change was studied further by using a waiting-time ($t_w$) protocol, in which $n_s$ is changed only temporarily during $t_w$, but $\Delta E_F>E_F$.  As a result of such a large perturbation, the system exhibits visible relaxations of $\sigma$, i.e. aging.\cite{Jan2006,Jan2007-1,Jan2007-2}  Interestingly, the non-Gaussian PDFs exhibit both history dependence and an evolution towards a Gaussian shape as the system ages and slowly approaches equilibrium, similar to the behavior of a great variety of out-of-equilibrium systems.

The power spectra and the higher order noise statistics of the non-Gaussian noise have been also investigated over a wide range of the applied bias voltage.  The data demonstrate that the non-Gaussian noise is observed in the regime of linear response, i.e. that it is not caused by the application of a high bias, but rather that it reflects the intrinsic out-of-equilibrium behavior of the 2DES.

In summary, we have established several new characteristics that the 2DES in Si has in common with a large class of both 2D and 3D out-of-equilibrium systems.  Thus this work provides additional strong evidence that many such universal features are robust manifestations of glassiness, regardless of the dimensionality of the system.  In addition, we have revealed some effects that are unique to Coulomb glasses.  Therefore, our findings should be helpful in the understanding of the complex behavior near the MIT in a variety of strongly correlated materials.

\acknowledgments
This work was supported by NSF DMR-0905843, the National High Magnetic Field Laboratory through NSF Cooperative Agreement DMR-0654118, and the State of Florida.

\end{document}